# In situ engineering hexagonal boron nitride in van der Waals heterostructures with selective SF$_6$ etching

Hitesh Agarwal[1], Antoine Reserbat-Plantey*[,1,2], David Barcons Ruiz[1], Karuppasamy Soundarapandian[1], Geng Li[1], Vahagn Mkhitaryan[1], Johann Osmond[1], Helena Lozano[1], Kenji Watanabe[3], Takashi Taniguchi[4], Petr Stepanov[1,5], Frank. H. L. Koppens*[,1,6], Roshan Krishna Kumar*[,1,7].

[1] ICFO - Institut de Ciències Fotòniques, The Barcelona Institute of Science and Technology, Av. Carl Friedrich Gauss 3, 08860 Castelldefels (Barcelona), Spain
[2] Université Côte d'Azur, CNRS, CRHEA, rue Bernard Grégory, 06560 Valbonne, France
[3] Research Center for Functional Materials, National Institute for Materials Science, Tsukuba 305-0044, Japan
[4] International Center for Materials Nanoarchitectonics, National Institute for Materials Science, Tsukuba 305-0044, Japan
[5] Department of Physics and Astronomy, University of Notre Dame, Notre Dame, Indiana 46556, United States
[6] ICREA - Institució Catalana de Recerca i Estudis Avancats, 08010 Barcelona, Spain
[7] Catalan Institute of Nanoscience and Nanotechnology (ICN2), Campus UAB, Bellaterra, 08193 Barcelona, Spain

*corresponding authors, roshan.krishna@icn2.cat, antoine.reserbat-plantey@cnrs.fr, frank.koppens@icfo.eu

**Van der Waals heterostructures are at the forefront in materials heterostructure engineering, offering the ultimate control in layer selectivity and capability to combine virtually any material. Hexagonal-boron nitride, the most commonly used dielectric material, has proven indispensable in this field, allowing the encapsulation of active 2D materials preserving their exceptional electronic quality. However, not all device applications require full encapsulation but rather require open surfaces, or even selective patterning of hBN layers. Here, we report on a procedure to engineer top hBN layers within van der Waals heterostructures while preserving the underlying active 2D layers. Using a soft selective SF$_6$ etching combined with a series of pre and post-etching treatments, we demonstrate that pristine surfaces can be exposed with atomic flatness while preserving the active layers' electronic quality. We benchmark our technique using graphene/hBN Hall bar devices. Using Raman spectroscopy combined with quantum transport, we show high quality can be preserved in etched regions by demonstrating low temperature carrier mobilities > 200,000 cm$^2$/Vs, ballistic transport probed through magnetic focusing, and intrinsic room temperature phonon-limited mobilities. Atomic force microscopy brooming and O$_2$ plasma cleaning are identified as key pre-etching steps to obtaining pristine open surfaces while preserving electronic quality. The technique provides a clean method for opening windows into mesoscopic van der Waals devices that can be used for local probe experiments, patterning top hBN in-situ, and exposing 2D layers to their environment for sensing applications.**

The synthesis of high-quality hexagonal-boron nitride[1] marked a turning point in two-dimensional materials research[2–4]. As an inert 2D crystal, it is an excellent dielectric material in 2D electronics[5], provides atomically flat surfaces with pristine interfaces[6], and protects active 2D layers from degrading atmospheric environments[7]. These properties enable engineering van der Waals heterostructures of the highest quality with unique functionalities envisioned for next-generation semiconductor technologies[8]. Additionally, they serve as powerful condensed matter simulators harboring physics spanning strongly correlated electron phenomena to topological physics[9]. In most van der Waals heterostructures, hBN typically encapsulates active 2D layers. The bottom hBN protects layers from rough substrates[10], while the top isolates it completely from its environment, pushing device quality to intrinsic limits[11]. However, certain applications and experiments require exposed surfaces. Hence, full encapsulation is not always desirable because it restricts access to active 2D layers.

From an applications perspective, exposed surfaces offer advantages and novel functionalities, such as tailoring light-matter interactions via patterned metasurfaces[12], leveraging 2D materials' surface sensitivity for biological and chemical sensing[13] and ensuring good electrical and magnetic contacts[14]. From a fundamental standpoint, exposed surfaces allow direct access to the underlying electron system using local probes. Some of the most powerful local spectroscopic probes, including scanning tunneling microscopy[15,16,17,18] (STM) and Angle-resolved photoemission spectroscopy[19,20] (ARPES), require exposed 2D layers achievable only through challenging heterostructure engineering. These techniques limit scanning areas and suffer from polymer contamination which may degrade sample quality. In all these applications and experiments, the bottom hBN remains crucial playing a major role in preserving electronic quality.

To overcome these challenges, we introduce an in-situ hBN patterning method for van der Waals heterostructures. We demonstrate that sulfur hexafluoride ($SF_6$) can selectively etch[21] the top hBN in encapsulated heterostructures, opening windows into small regions of the device (Fig. 1) that serve as access points to the underlying 2D layers. Using graphene encapsulated with hexagonal-boron nitride, we track the quality of selectively etched regions through atomic force microscopy (AFM), Raman spectroscopy and quantum transport measurements, demonstrating that high electronic quality remains intact, limited by atmospheric conditions. The methodology is sketched in Figure. 1-a-h. Starting with fully encapsulated hBN/graphene/hBN devices with electrical contacts (Fig. 1a), we pattern etching masks (Fig. 1b) with PMMA. Following, a series of cleaning steps including $O_2$ plasma etching and AFM brooming (Fig. 1c-e) prepares the surface for $SF_6$ etching of hBN (Fig. 1f).

$SF_6$ has proven highly effective in graphene nanofabrication due to its ability to selectively etch hBN without damaging graphene, enabling controlled exposure of clean graphene edges [22]. When used as an etchant for hBN, the energetic plasma breaks $SF_6$ molecules, generating reactive fluorine radicals. These fluorine atoms react with the boron in the BN lattice forming volatile boron trifluoride ($BF_3$), while the nitrogen is released as molecular nitrogen[21] (N). In a simplified overall reaction, one may write: $2BN(s) + 6F \rightarrow 2BF_3(g) + N_2(g)$ where (s) and (g) refer to solid and gas phase respectively. In the plasma process, fluorine atoms come from $SF_6$, which is reduced to $SF_4$ or other sulfur-fluoride species. The main etch products are boron trifluoride ($BF_3$), removed as a volatile gas, and Nitrogen ($N_2$) released into the atmosphere. Because of its excellent selectivity for hBN over graphene, it is commonly used to create low-resistance electrical contacts[22]. Here, we use it instead to directly pattern the top encapsulating hBN layers, selectively removing specific areas that expose the underlying graphene (Fig. 1) while preserving the bottom hBN substrate and maintaining graphene's high electronic quality.

In this work, we use a recipe inspired by previous works[21,22]. However, we introduce additional pre-etching steps and refine the etching procedure to optimize graphene's electronic quality (see Supplementary Section 1), making it suitable for various applications and experiments. While $SF_6$ etching stops at graphene, over-etching may still be detrimental, leading to fluorination or doping inhomogeneities. Calibrating etching time and recipes across different Reactive Ion Etchers (RIE) is crucial to ensure selective hBN removal without excessive graphene fluorination. Generally, a stable plasma should be maintained with the highest chamber pressures but lowest radio frequency (RF) powers. Higher pressures reduce precision but enable slower, more controlled hBN removal, essential for preserving graphene's electronic quality. We perform systematic calibration using reference regions on hBN layers and use optical microscopy to track the hBN removal during sequential etching steps until complete. The calibrated recipe can then be applied to the active region. We characterize our devices in the etched regions before and after exposure of graphene to benchmark its electronic quality and the importance of the steps employed in **b – g**.

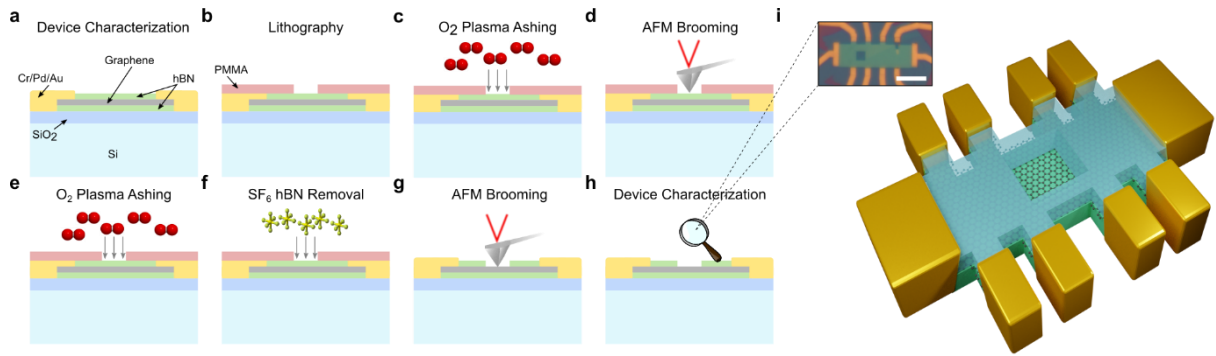

**Figure. 1. Methods for in-situ engineering top hexagonal-boron nitride in van der Waals heterostructures**. **a – h**, schematics of the work flow presented in this work. The device schematic is sketched in **a**, showing a graphene (grey) encapsulated with hBN (green), contacted with Chromium, platinum, gold (Cr/Pd/Au) and fabricated on a Silicon dioxide ($SiO_2$) on Silicon (Si) substrate which was used as an electrostatic gate. A series of lithography **(b)**, $O_2$ plasma ashing **(c,e)** and dry etching **(f)** is performed. At specific steps, AFM brooming is employed **(d,g)** to remove contaminants and preserve electronic quality. (**i**) optical image of our device D1 after a window has been opened. Scale bar is 10 μm.

To develop the selective etching recipe, we studied graphene-based/hBN heterostructures. This included two monolayer graphene devices encapsulated with hBN (D1/2) and a small angle twisted bilayer graphene device (D3). All devices were fabricated using standard methods (see methods). The process involved heterostructure assembly, followed by nanofabrication of mesoscopic devices with Hall bar geometries (Fig. 1i). Low-temperature (10 K) quantum transport measurements characterized the devices, revealing excellent electronic quality, including size-limited mobilities[23,24] (see discussions below). All devices were studied before and after etching using AFM. However, Raman spectroscopy and quantum transport measurements post-etching were performed exclusively on monolayer graphene/hBN heterostructures because of its well-known electronic properties which enabled proper calibration.

Following low-temperature device characterization, we applied our selective etching method. Optimized recipes, discussed in Methods and Supplementary Section 1, were tested on all devices. Here, we present results on hBN/Graphene/hBN devices, where three windows were opened on two devices—D1 (Fig. 2a) and D2 (Fig. 2c)—demonstrating the importance of pre- and post-etching steps. The first is a small square on the device's left side (Fig. 2a). For this window (D1_window1), we follow steps Fig. 1b-f, directly etching the device after patterning the PMMA mask. The second window (D1_window2) extends the entire device width. In this case, additional pre-etching steps were used, where the top hBN surface was first cleaned using AFM brooming[25,26,27] to remove polymer residues from the targeted region (Fig. 1d-e), followed by $O_2$ plasma cleaning. The third window in D2 (D2_window3) followed the same procedure, but was additionally annealed under ultra-high vacuum post-etching.

All three windows were characterized using AFM. Fig. 2a plots a topographic map of D1 after etching windows 1 and 2. The two regions exhibit notable differences. D1_window1 shows significant surface roughness resembling contaminants accumulating on the surface of graphene. In contrast, the D1_window2 appears much smoother with a root square mean roughness ($R_q$) below 1 nm. Fig. 2b further illustrates this by plotting the height profile (*t*) as a function of x,y spatial co-ordinates for a 1 x 1 μm$^2$ areas. We attribute these differences to the crucial importance of AFM brooming. Without brooming, any surface contamination on the top hBN falls onto the graphene. With brooming the top hBN can be cleaned enabling selective etching that exposes the graphene with a pristine surface. The residual surface roughness tells us that some contamination still remained, possibly due to hydrocarbon absorbates or PMMA residues. Following the same procedure in a third device (D3_window4), we achieved atomically flat areas ($R_q$ = 0.2 nm) over 1 x 1 μm$^2$ areas (see Supplementary Section 2). With additional post-etching annealing in D2_window3, even smoother surfaces were achieved. This included atomically flat areas over 2 x 2 μm$^2$, meeting the requirements of sensitive scanning probe experiments (Fig. 2c).

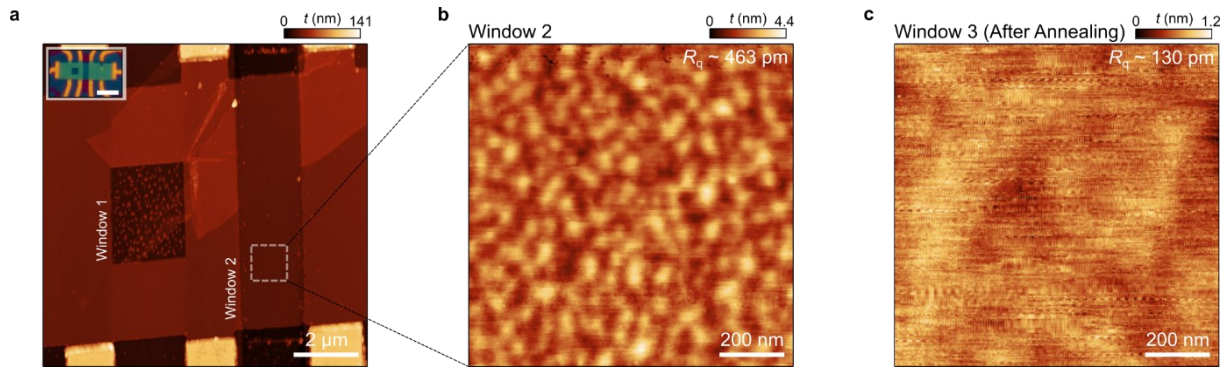

**Figure. 2. Atomic force microscopy measurements of etched regions**. **a,** an AFM image of device 1 in the regions where D1_window1 and D1_window2 have been opened (see white labels). The gold regions with the largest height (*t*) correspond to the electrical contacts of the device. The inset displays an optical image of the device. The scale bar is 2 μm. **b**, Zoomed spatial map of the height profile (*t*) in D1_window 2. The area is indicated by the white dashed box in **a**. **c**, a 1 x 1 μm$^2$ area of D2_window3 after annealing.

While AFM measurements suggest clean exposed surfaces can be achieved, the chemical procedure may fluorinate graphene or degrade its crystalline quality, which AFM alone cannot easily detect. Therefore, we performed micro-Raman spectroscopy to characterize the exposed graphene's quality further, compare it with the fully encapsulated graphene, and evaluate the impact of etching. We focus on the $E_{2g}$ Raman active modes, leading to the G band[28] at around 1582 cm$^{-1}$, and for hBN[29] at 1362 cm$^{-1}$. Notably, we observed the 2D band in our exfoliated graphene monolayer sample, with intensity more than 5x times larger than the G band and with a Lorentzian profile, an indicator of weakly doped monolayer graphene[30] (Supplementary Information 3). At no point did we detect the emergence of the D or D' bands from graphene[31], even after opening the hBN window (Fig. 3b). This absence of the defect bands suggests good preservation of the crystalline structure after etching and therefore negligible fluorination.

We first performed hyperspectral mapping of the sample (Fig. 3a), wherein the laser spot (500 nm diameter) was scanned over the device, and a spectrum was recorded at each point. We identified the etched zones —D1_window1/2 – by plotting the hBN $E_{2g}$ phonon peak area ($A_{hBN}$) centred at 1362 cm$^{-1}$ (Fig. 3b) defined as $A_{hBN} = \pi/2\,(I_{hBN}\Gamma_{hBN})$, where $I_{hBN}$ and $\Gamma_{hBN}$ refer to the peak intensity and peak width respectively. Removing hBN material alters the optical planar cavity formed by the heterostructures on the SiO$_2$ dielectric and the back silicon mirror[32]. The interferences, governed by the local optical gain dependent on different layers' thicknesses and refractive indices, affect the pump laser (at 532 nm) and the Raman scattered light[33]. Consequently, the net interference pattern can be complex[34]. To better observe the effect of removing the hBN top layer, we normalized the hBN peak area by the G band area, assuming it remains constant during etching. This assumption is supported by the absence of the D band post etching, suggesting good preservation of sp² carbon bonds and confirming graphene's crystalline integrity[34]. We report a ~ 35% reduction in the hBN area signal in the window zone (Fig. 3b). Interestingly, a 55% decrease was expected, given the top hBN thickness of 17.5 nm and the bottom of 14 nm (measured *via* AFM). This discrepancy may result from normalizing by the G band area, which does not fully eliminate the Raman interference effect, as there is still a small shift in the scattered light wavelength between the two peaks.

To further analyse the quality of graphene after the etching, we plotted the correlation between the G and 2D band positions[35] (Fig. 3c), isolating clusters of points corresponding to the D1_window2 area (pink) and the fully encapsulated zone (blue). The two rectangular zones are shown in Fig. 3a. Although both clusters are close, we observe a systematic shift typically linked to doping differences between regions. This shift aligns with a local change in the Fermi level due to contaminant absorption directly on graphene or fluorination[35]. Following the framework analysis of previous works [10,35], which shows doping around $0.1 \times 10^{12}$ cm$^2$/Vs, consistent with our low temperature quantum transport measurements (see Supplementary Section 5). In contrast, encapsulated graphene remains protected from the environment. In both cases, data clusters elongate along the strain axis, indicating the strain distribution is not perfectly uniform across the scanned area[10]. Interestingly, elongation is more pronounced in the etched window, suggesting that removing the top encapsulant caused a local strain redistribution within the monolayer.

We then performed an additional step called AFM brooming, where an AFM tip in scanning contact mode cleans the exposed surface, acting as a nanoscale broom. We repeated the same Raman analysis after brooming (Fig. 3d) and observed that the data clusters almost perfectly overlap, indicating efficient removal of contaminants and a nearly identical Fermi level in both etched and encapsulated areas. Further Raman correlation analysis, comparing G and 2D linewidths (Fig. 3e–f), provides insight into strain distribution at the nanoscale[10] . Here, we observe a clear difference in strain distribution between the fully encapsulated pristine heterostructure and the etched window. The higher center position of the data cluster for the etched window suggests more inhomogeneous strain distribution at the nanoscale, varying over a smaller length scale than in the fully encapsulated case. At first glance, this may seem unexpected since encapsulation between hBN layers initially established graphene's strain distribution. However, removing the top hBN layer creates mechanical asymmetry and relaxes boundary conditions that stabilize strain distribution[36,37]. As a result, graphene undergoes local strain redistribution when the top encapsulant is removed, following strain transfer mechanisms similar to those in thin-film structures[38]. Notably, strain distribution does not change significantly after brooming, but the two clusters in Fig. 3f move closer due to reduced and more uniform doping.

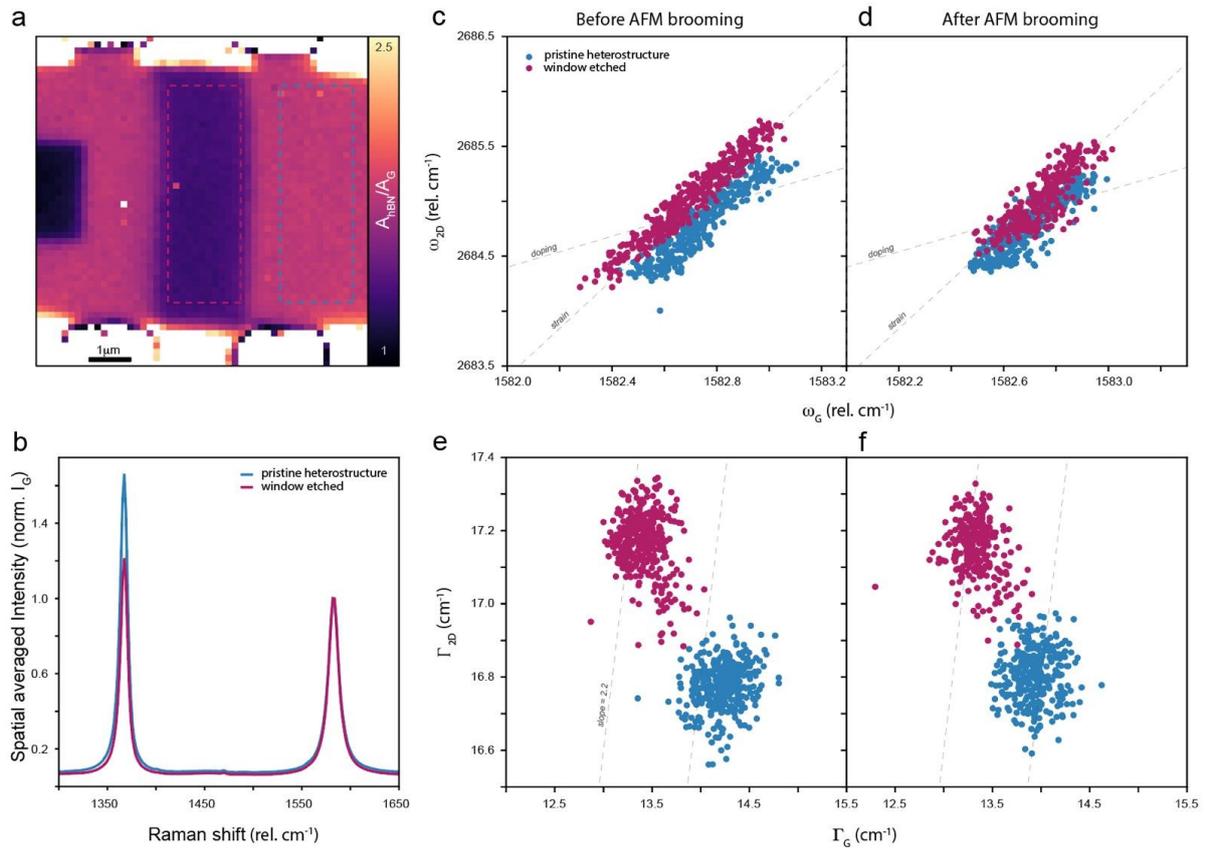

**Figure. 3 Investigation of window opening and surface brooming effects on graphene via micro-Raman spectroscopy. a**: Hyperspectral Raman map of the sample displaying the amplitude of the hBN $E_{2g}$ peak normalized by the amplitude of the graphene G peak, which mitigates interference effects. The zones highlighted in pink including the left square area, correspond to regions where the top hBN layer has been removed (etched windows). **b**: Spatially averaged Raman spectra of the $E_{2g}$ modes for hBN and the graphene G band. The spatial averages correspond to the pink (etched windows, D1_window2) and blue (fully encapsulated heterostructure) regions indicated in panel **a**. **c-d**: Correlation plots of the positions of the graphene G and 2D bands before (**c**) and after (**d**) AFM brooming. The pink and blue data points correspond to the etched windows and fully encapsulated areas shown in panel **a**, respectively. The black dashed lines represent theoretical predictions of ($\omega_G$, $\omega_{2D}$) for graphene under constant doping and uniform strain[35]. **e-f**: Similar correlation plots as in panels **c** and **d**, but showing the relationships between the linewidths of the G and 2D peaks. The dashed lines, with a slope of 2.2, correspond to the expected trend for strain-induced broadening of the Raman peaks[10], illustrating the increased broadening due to strain variations within the laser spot area.

Next, we characterized exposed regions using low-temperature quantum transport, comparing its quality with previous measurements before etching and in fully encapsulated regions. The blue curve in Fig. 4a plots the mobility extracted from the Drude conductivity measured at 10 K before etching. It shows mobility's larger than 400,000 cm²/Vs over the entire doping range. In pristine graphene heterostructures, the channel is so clean that carriers do not scatter until reaching device edges, resulting in a mobility limited by the device width[23,24]. The green solid line in Fig. 4a plots the width-limited mobility corresponding to our devices ($w$ = 8 μm). For hole doping ($n < 0$), mobilities reach this limit demonstrating excellent electronic quality. For electron doping mobility is slightly less, potentially due to impurities in the hBN substrate. Because of the excellent electronic quality, our devices exhibited clear signatures of ballistic transport, including magnetic focusing[39–41] and negative bend resistance[7].

The AFM and Raman data from D1_window2 are particularly promising, suggesting high-quality graphene can be preserved even after removing top hBN provided AFM brooming[26,27] is employed. To benchmark electronic quality, we performed low temperature quantum magneto-transport measurements, comparing device characteristics before and after opening the windows. The pink data in Fig. 4a plots the mobility of our device measured at 10 K in the same region, using the same contact pairs, before the window was opened (inset of Fig. 1a). While the mobility decreases for all the doping, it still remains relatively high > 200,000 cm$^2$/Vs, competitive with the state-of-the-art reports of graphene on hBN without encapsulation[6,42]. Even with pristine etching that avoids graphene fluorination, some degradation may be expected because the graphene is now exposed to atmospheric conditions and may be sensitive to absorbates.

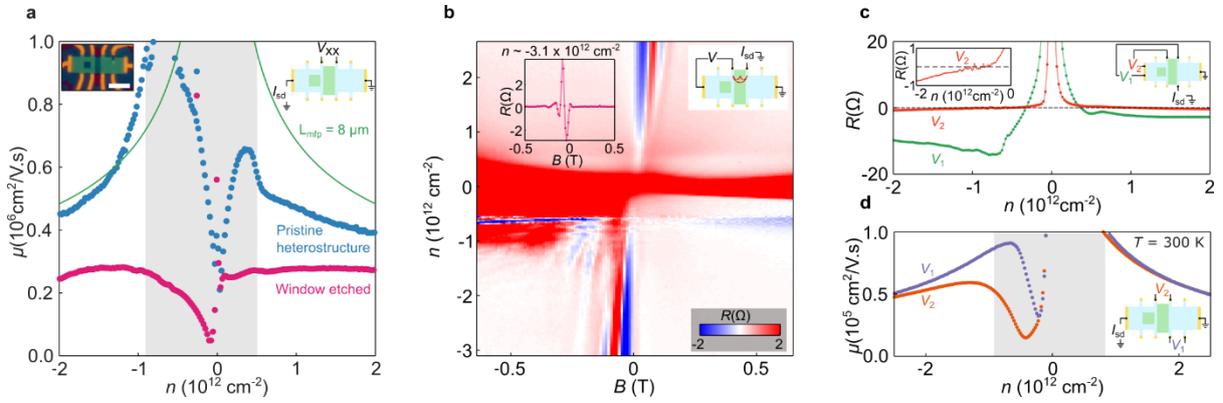

**Figure 4. Quantum transport measurements in etched region a**, mobility ($\mu$) as a function of carrier doping ($n$) measured at 10 K in the same region of the device before (blue circles) and after (pink circles) etching D1_window1 and D2_window2. The green solid line traces the device width limited mobility. The top right inset sketches the measurement geometry. **b,** Magnetic focusing resistance $R_{TMF}$ plotted as a function of carrier doping ($n$) and magnetic field ($B$) at 10 K. Top left inset plots a line cut for carrier density $n$ = -3.1 x 10$^{12}$ cm$^{-2}$. Top right inset sketches the measurement geometry. **c** Bend resistance $R_B$ plotted as a function of carrier doping $n$ at 10 K for two geometries $V_1$ and $V_2$. Top right inset sketches the measurement geometries. **d,** $\mu$ ($n$) measured around D1_window2 (orange) and in the fully encapsulated region (purple) at 300 K.

To further access device quality, we performed magnetic focusing experiments[39–41] (inset of Fig. 4b). These measurements serve as an excellent characterization tool, tracking ballistic trajectories of semi-classical quasi-particles exhibiting Lorenz-like motion. The measurement geometry is sketched in the inset of Fig. 4b. In magnetic focusing experiments, ballistic carriers injected at side contacts are curved by finite magnetic fields and collected at adjacent electrodes. For specific values of magnetic field, when the semi-classical cyclotron radius is commensurate with contact spacing, resistance peaks appear. We choose a geometry that tracks ballistic properties through the etched region. An example of the resonances can be seen clearly in the inset of Fig. 4b showing an oscillatory structure appearing for one sign of the magnetic field. Fig. 4b plots the magnetic focusing resistance $R_{TMF}$ as a function of magnetic field ($B$) and carrier density ($n$). Notably, strong magnetic focusing resonances can be observed for electron and hole doping. For hole doping, higher-order resonances corresponding to multiple reflections from device edges are visible up to $p$ = 5. The observation of magnetic focusing demonstrates the excellent quality of our devices even after removal of the top hBN. To our knowledge, this is the first demonstration of ballistic transport over micron-length scales in single-sided graphene encapsulation. To further characterize the ballistic properties, we performed bend resistance measurements. The geometries are sketched in the inset of Fig. 4c. For $V_1$ we observe a strong negative bend resistance appears with doping indicating carriers propagate ballistically across the device over 8 µm. In $V_2$ the negative response was strongly suppressed, likely due to trajectories passing through the etched region where the mobility is lower (Fig. 4a). Nonetheless, ballistic carriers remain detectable (Fig. 4c inset). The strong negative bend resistance in $V_1$ confirms that unetched regions retain high electronic quality after $SF_6$ treatment. Similar observations were made near D1_window1. Although mobility degradation was more severe due to surface contamination, high-order magnetic focusing features remained visible, corresponding to trajectories avoiding the window region (see Supplementary Section 4). This further demonstrates that high electronic quality is preserved in unetched regions.

Fig. 4a shows that carrier mobility does not degrade significantly at 10 K and remains high across the doping range. For device applications, assessing quality at room temperature is also crucial. Fig. 4d plots mobility measured in the etched region (D1_window2) and pristine fully encapsulated regions at 300 K. At these temperatures, mobility appears largely unaffected by opening a window in the hBN, remaining nearly identical for hole doping. We attribute this behavior to dominant phonon contributions, which limit graphene's intrinsic mobility. In other words, at room temperature, etched regions approach the intrinsic mobility limits of graphene encapsulated with hBN[11] with competitive values > 50,000 $cm^2$/Vs.

**Discussion**

Our measurements show that clean in-situ patterning of top hBN in van der Waals heterostructures can be made without significantly degrading the electronic quality of underlying 2D layers. AFM measurements confirm that pristine surfaces can be opened if AFM brooming is performed. Our experiments showed that this step is essential in obtaining clean surfaces after etching with < 1 nm surface roughness (Fig. 2a-b). However, some variability in achievable surface roughness on the sub nanometer scale remains to be understood. The cleanest windows were observed in D2_window3 (Fig.2c) and D3_window4 (supplementary section 2), showing regions of atomically flat surfaces ( < 200 pm). This enhanced quality compared to D1_window2 may be due to several factors. First, molecular adsorbates may contribute to the surface roughness[43]. Second, the AFM brooming conditions may require further optimization, depending on the number of pre-etching steps involving PMMA deposition. In D1, polymer masks were deposited and washed twice for opening D1_window1 and then D1_window2. In contrast, D2_window3 and D3_window 4 underwent one etching procedure, uniquely defined on different devices. Furthermore, the PMMA mask was not removed after etching in D3_window4. In summary, D2/D3 generally faced less exposure to PMMA contamination than D1. This step may be important for achieving the atomically pristine surfaces, as washing in solvents can lead to nanoscale contaminants adsorbing onto the window region. Nonetheless, residual PPMA can still be removed through high-temperature annealing. This was evident in D2_window3 which showed a vast improvement in surface roughness post-annealing (see Supplementary Section 7). Further work is needed to optimize processing steps for reproducing pristine atomic surfaces.

In some applications, selective etching may be needed before depositing contacts, such as when 2D contacts to exposed graphene are required. In this case, Raman spectroscopy can characterize graphene quality due to its sensitivity to fluorination, serving as an initial screening of the active 2D layer before contact deposition. Strong fluorination is typically detected through the appearance of D' peaks[44,45]. However, previous studies required long exposure times to resolve Raman signatures.

Our etching recipes use significantly shorter times, lasting minutes rather than hours. Thus, it is unsurprising that excess fluorination is not observed. However, low-temperature quantum transport measurements show slight quality degradation. When plotting resistivity as a function of carrier doping (see Supplementary Section 5), we notice peak splitting at the Dirac point, suggesting doping inhomogeneities in the sample. This may result from slight fluorination undetectable by Raman signals or surface contaminants. Notably, brooming the device surface after etching seemingly restored pristine quality (Fig. 3d/f), indicating degradation likely originates from surface contamination.

Finally, we note that over-etching can degrade graphene´s electronic quality. This was evident in D2_window3, where we intentionally over-etched for 2 minutes. Raman and quantum transport measurements on this sample showed strong doping inhomogeneities post-etching (see Supplementary Section 6). This experiment highlights the need for careful calibration of etching procedures and demonstrates that room-temperature Raman spectroscopy effectively probes surface quality in exposed graphene, even without the presence of D' bands. However, the surface can be further treated through high-temperature annealing. This is evident in Raman spectroscopy measurements performed on D3 before and after annealing (see Supplementary Section 6), showing that doping and strain profiles change post-annealing. Additionally, annealing at moderate temperatures (250°C in an argon-hydrogen mixture with 10% $H_2$) may help reverse fluorination effects[44].

**Conclusion**

Our experiments outline a new technique for engineering the top encapsulating hBN in van der Waals heterostructures. We demonstrate the technique can be used to open windows into graphene encapsulated with hexagonal-boron nitride, while preserving the electronic quality of the 2D surface. Aside from hBN, preliminary experiments on selectively etching transition-metal dichalcogenides (see Supplementary Section 8), show that high quality of underlying graphene can also be preserved, highlighting the broader applicability to graphene encapsulated with 2D semiconductors[46]. The technique has strong prospects in scanning probe microscopy, enabling the possibility to access fully encapsulated quantum transport devices to bridge the gap in understanding between global and local measurements. With further optimization, the procedure offers a more controllable method for obtaining pristine surfaces compared to the stack-and-flip method currently employed[20]. Cryogenic techniques may further enhance controllability and resolution in advanced nano-patterned structures[47]. It may also be used in Scattering near-field optical microscopy (SNOM) experiments, enhancing the light-matter coupling between the tip and the underlying substrate[48]. From an application perspective, the capability to pattern the top hBN offers exciting opportunities. Aside from chemical sensing, it offers unique directions in 2D meta-materials[49] including engineering of polaritonic launches, excitonic landscapes, advanced[50] and artificial superlattice structures[51,52].

**Methods**

**Device Fabrication:** The samples are fabricated using typical methods in heterostructure assembly. Typically, a thin hBN flake (~10-15 nm) is picked using hot-pick up technique [22,53] using a polypropylene carbonate (PC) film on a polydimethylsiloxane (PDMS) stamp at 90º C. This hBN flake is then later used to pick up the graphene monolayers, mechanically exfoliated on $Si^{++}/SiO_2$ (285 nm) from highly oriented pyrolytic graphite, and pre-characterized using optical microscopy, and Raman spectroscopy [28]. In D3, twisted graphene was assembled between hBN layers and an additional graphite layer was picked up after the bottom hBN. Finally, the stack is used to pick up a last layer of hBN and later dropped on a pre-patterned marker chip of $Si^{++}/SiO_2$ (285 nm) at 180 º C, squeezing out the bubbles, and impurities as previously reported. The stack is then shaped into a Hall bar geometry using $SF_6$ plasma, $O_2$ plasma to etch top hBN, and graphene respectively, and further metalized using 3/15/30 nm of Cr/Pd/Au. The thickness of hBN flakes is identified by their colour shading through optical contrast in a microscope, and consequently via AFM when assembled in the heterostructure. The heights can be distinguished from steps in the height profiles induced by exposed graphene edges.

**$SF_6$ etching rates:** The etching was performed in an OXFORD or SAMCO reactive ion etcher (RIE) (see Supplementary Information 1 for detailed recipe). For D1_window1, D1_window2, and D2_Window3, etching were done in OXFORD RIE with hBN etch rate of ~ 7 nm/min. For D3_window4, etching was done in SAMCO RIE with hBN etch rate of ~ 5 nm/min. Before etching, the RIE chamber was pre-conditioned for 15 minutes with the same $SF_6$ recipe.

**Raman Spectroscopy:** Raman spectroscopy measurements are performed at room temperature using a 532 nm laser with optical power of 0.5mW, focused on a 500 nm spot.

**Quantum transport:** Low temperature quantum transport measurements were performed on an Advanced Research Systems 4 K Cryostat. Measurements were performed using standard lock-in techniques (SR 860) in constant current mode sourcing a small AC current (< 100 nA) while measuring the four-probe voltage. For magnetic field measurements, a 1 T electromagnet was used (GMW associates).

## Acknowledgements

We would like to thank Carmen Rubio-Verdú (ICFO), Tymofiy Khodkov (ICFO), Lene Gammelgaard (DTU), Juan Sierra, Patricia Aguilar, Sergio Valenzuela (ICN2) and Qian Yang (University of Manchester) for important insights and useful discussions. We further thank Matteo Ceccanti for making the illustration presented in Figure. 1. H.A. acknowledges funding from the European Union's Horizon 2020 research and innovation programme under Marie Skłodowska-Curie grant agreement no. 665884. A.R-P acknowledges funding from ANR JCJC NEAR-2D and Welcome Package Idex (UniCA) as well as AAP Tremplin Complex 2023 "2DNEUROTWIST" (ANR-15-IDEX-01). R.K.K. acknowledges funding by MCIN/AEI/ 10.13039/501100011033 and by the "European Union NextGenerationEU/PRTR" PCI2021-122020-2A within the FLAG-ERA grant [PhotoTBG], by ICFO, RWTH Aachen and ETHZ/Department of Physics. R. K. K also acknowledges support from the Ramon y Cajal Grant RYC2022-036118-I funded by MICIU/AEI/10.13039/501100011033 and by "ESF+". F.H.L.K. acknowledges support from the ERC TOPONANOP under grant agreement no. 726001, the Gordon and Betty Moore Foundation through Grant GBMF12212, and the Government of Spain (FIS2016-81044; PID2019-106875GB-100; Severo Ochoa CEX2019-000910-S [MCIN/ AEI/10.13039/501100011033], PCI2021-122020-2A, and PDC2022-133844-100 funded by MCIN/AEI/10.13039/501100011033). This work was also supported by the European Union NextGenerationEU/PRTR (PRTR-C17.I1) and EXQIRAL 101131579, Fundació Cellex, Fundació Mir-Puig, and Generalitat de Catalunya (CERCA, AGAUR, 2021 SGR 014431656). Views and opinions expressed are those of the author(s) only and do not necessarily reflect those of the European Union Research Executive Agency. Neither the European Union nor the granting authority can be held responsible for them. Additionally, the research leading to these results received funding from the European Union's Horizon 2020 under grant agreements no. 881603 (Graphene flagship Core3) and 820378 (Quantum flagship). This material is based upon work supported by the Air Force Office of Scientific Research under award number FA8655-23-1-7047. Any opinions, findings, conclusions, or recommendations expressed in this material are those of the author(s) and do not necessarily reflect the views of the United States Air Force.

**Supplementary Information 1 – Etching recipe for etching hBN**

To pattern the top hexagonal-boron nitride (hBN) of graphene encapsulated in hBN we employ the following steps. We first spin-coat a single layer of PMMA (PMMA 950) and open a window via standard e-beam lithography in the area of interest. We then carefully clean the opened area of interest by pushing the dirt (PMMA residue or dirt) using the contact mode AFM brooming technique. The surface is then treated with a low power mild $O_2$/Ar plasma (Recipe 7 - Moorefield RIE, L106, RF Power = 1 W, Time = 3 min, DC self-bias ∼ 55 V) to activate the surface and create free radicals on the surface. We then Dry etch the sample using pure $SF_6$ in small steps (Recipe – 'SF$_6$ selective hBN NOE – Oxford RIE', Pressure = 100 mTorr, $SF_6$ = 40 SCCM, RF Power = 20 W, DC Bias ∼ 13 V) until top hBN is not visible in an optical microscope. It is crucial to precondition the etching chamber with the same recipe for 15∼20 mins, and pump the load-lock until 5 x 10$^{-6}$ Torr before starting the etching. Following this, the PMMA mask may be removed by leaving the sample in acetone for overnight lift-off. This step is not always essential, and it may actually be beneficial to leave the PMMA to reduce the risk of contamination in the open surfaces.

**Supplementary Information 2 – Window opening in Device 3**

In Device 1, we found atomic force microscopy (AFM) brooming makes a significant difference in obtaining clean open surfaces. This was made evident by the vast improvements between D1_window1 and D1_window2 (Fig. 2a in main text). In D1_window2, high electronic quality could be achieved but a surface roughness of around 463 pm per 1 μm$^2$ prevailed. While additional annealing can improve the surface roughness significantly (as seen in Fig. 2c of main text), this step is not always desirable as it risks contact degradation due to the harsh temperature conditions. Here, we show smoother surfaces without high-temperature annealing can be achieved using a third device (D3). The device consists of a twisted bilayer graphene encapsulated with hBN placed on top of a bottom graphite gate (Inset of Fig. S2a). An AFM image of the window region is shown in Figure S2a and a zoom plotted in Fig. S2b. Notably, we found much lower surface roughness around 164 pm could be achieved in this device. We believe the high-quality surface was achieved because the device underwent less lithographic steps which reduces the amount of PMMA contamination. Moreover, we did not wash the PMMA mask after $SF_6$ etching which may be a crucial step in reducing contamination on the open surface.

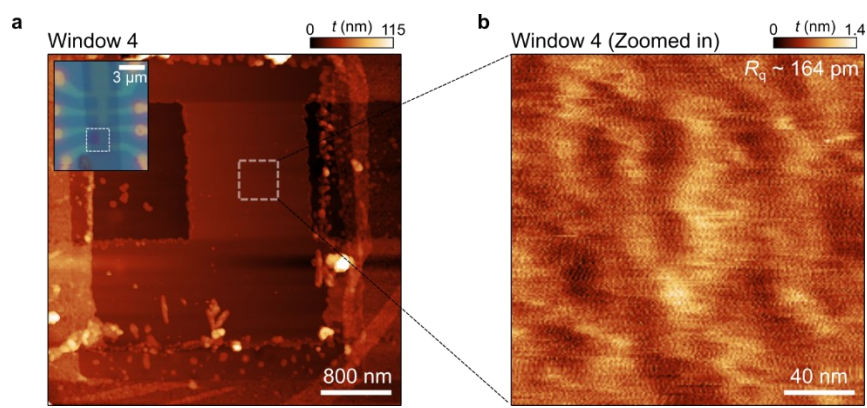

**Figure. S2 Atomic force microscopy in Device 3 a**, AFM image of Device 3 after etching. Inset: optical image of the device. The scan area is indicated by the white dashed box on the inset. **b,** A zoomed higher resolution AFM image (indicated on **a**).

**Supplementary Section 3 – Full Raman spectrum on graphene encapsulated with hBN**

In figure. 3 of the main text, we present Raman analysis performed only on the G peak of graphene and hBN phonon resonance. Here we present the whole spectrum for pristine and etched regions of D1_window2. Both show the behaviour of ultra-high quality graphene including a prominent 2D peak, and complete absence of a D' band observed in defective graphene[1].

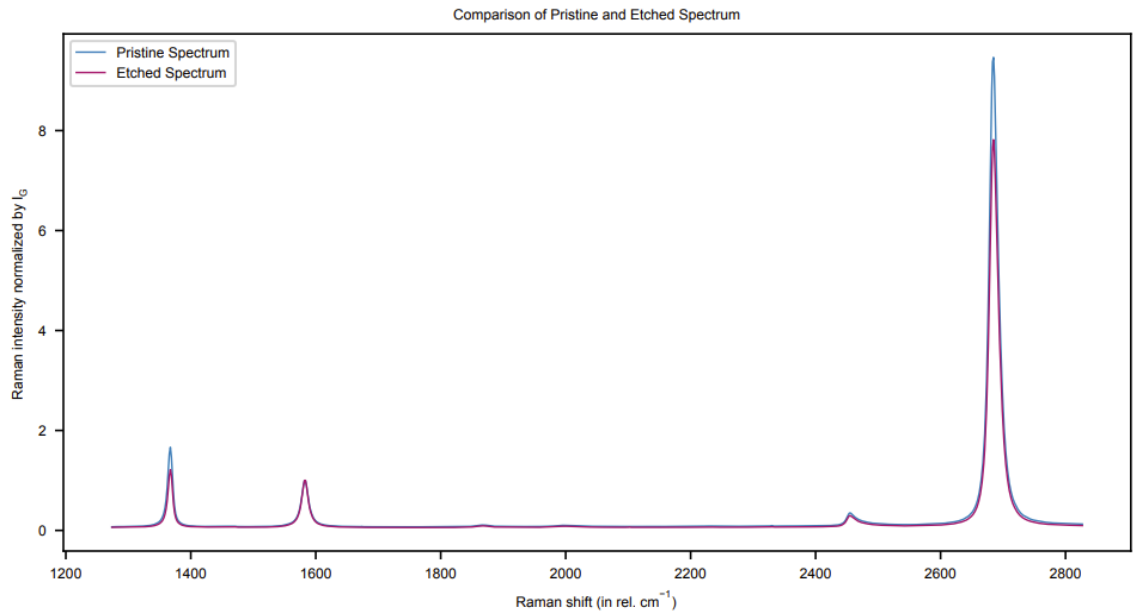

**Figure. S3 Raman spectrum in Device 2.** The blue and pink data plots the Raman spectrum in pristine and etched regions respectively.

**Supplementary Section 4 – Quantum transport around D1_window1**

Here we present quantum transport data around D1_window1, where significant PMMA residue falls into the graphene. Figure. S4a plots the carrier mobility in the window region before and after etching. It shows similar behaviour to the rectangle D1_window2, except a slightly stronger mobility degradation. However, this measurement includes also pristine regions and hence does not represent the true electronic quality of the window. For this, we performed ballistic transport measurements. Specifically, we used magnetic focusing[2–4] and negative bend resistance[5] to track ballistic trajectories passing through the window region. The blue and pink data set in Fig. S4b plots the magnetic focusing resistance $R$ as a function of magnetic-field ($B$) for a fixed carrier doping of holes, before and after the etching. Before the etching, a strong oscillatory structure can be observed with all the usual magnetic focusing resonances observed up to an order of $p$ = 3, corresponding to cyclotron trajectories following multiple reflections along device edges (inset of Fig. S4b) – the usual behavior of ballistic graphene heterostructures. After etching, we resolve the higher-order reflections but the lowest order $p$ = 1 are suppressed. This is not surprising considering the geometry of the window. As sketched in the inset, the lowest order trajectories with largest cyclotron radius pass directly through the window region where the mobility is lower. The first negative peak corresponds to those that must pass through longer distances of the window region[6]. Its strong suppression tells us the quality degrades. To further understand the behavior of ballistic electrons in the window region, we performed zero-field bend resistance measurements (inset of Fig. S4d). In this geometry negative bend resistance corresponds to trajectories that propagate directly across the device collected at contacts. We measured two geometries ($V_1$ and $V_2$). The former corresponds to trajectories that pass outside the window region, exhibiting a strong negative response for hole doping where the mobility is maximum. In contrast $V_2$ exhibits very little negative response indicating the ballistic transport is suppressed due to lower electronic quality, consistent with the magnetic focusing data. Nonetheless, magnetic focusing features could be consistently observed for all doping (Fig. S4c) up to orders of $p$ =5 over length scales of 5 µm. This demonstrates that the electronic quality of the encapsulated regions remains pristine.

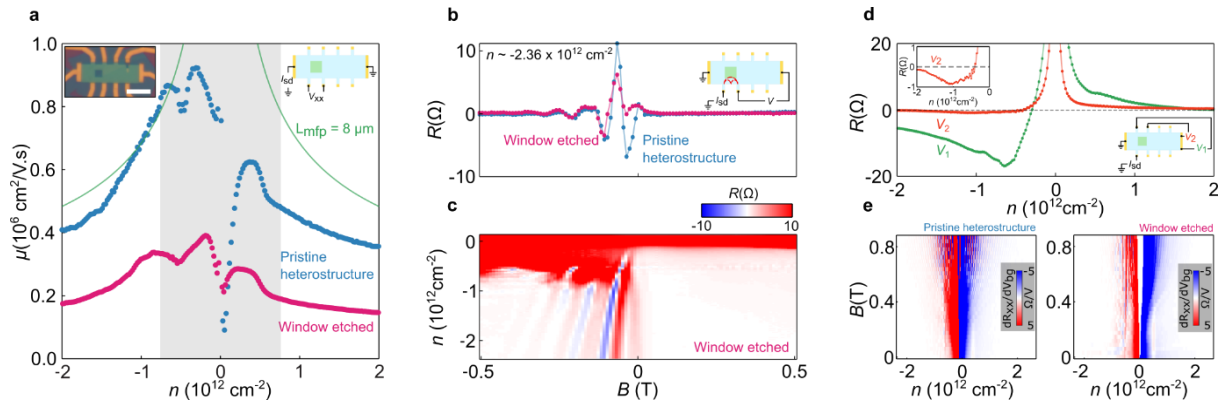

**Figure. S4 Quantum transport measurements in etched region a**, mobility (µ) as a function of carrier doping ($n$) measured at 10 K in the same region of the device before (blue circles) and after (pink circles) etching D1_window1. The green solid line traces the device width limited mobility. The top right inset sketches the measurement geometry, and the top left inset is the optical picture of the device with scale bar 10 µm. **b,** Magnetic focusing resistance $R_{TMF}$ plotted as a function of magnetic field ($B$) at 10 K. The data is plotted before and after etching for a fixed carrier doping of $n$ ~ -2.36 x $10^{12}$ cm$^{-2}$. Top right inset sketches the measurement geometry. **c,** $R_{TMF}(B,n)$ measured in the same geometry sketched in **b**. **d**, Bend resistance $R_B$ plotted as a function of carrier doping ($n$) at 10 K for two geometries $V_1$ and $V_2$. Bottom right inset sketches the measurement geometries. **e,** $R_{xx}$ as a function of magnetic field measured using the geometry sketched in **a**, before and after opening the window. It shows low-field Shubnikov de-Haas oscillations can still be observed.

**Supplementary Section 5 – Inhomogeneities induced by weak fluorination observed in low-temperature quantum transport measurements**

Figure. 4a in the main text shows that the graphene channel preserves its electronic quality even after exposing it using our selective etching recipe. Here we present the raw data measured in D1_window2 to help understand where mobility degradations come from. Figure. S5a plots the resistivity $\rho_{xx}$ as a function of the carrier doping ($n$) obtained via Hall effect measurements. The blue and pink data-set plots the behaviour before (pristine heterostructure) and after window etching. We notice two things. First of all, the resistivity increases, seen clearer on a logarithmic scale (Fig. S5b). Second, there is a slight shift of the Dirac point. This shift tells us that the sample attains some intrinsic doping. The broadening of the Dirac point, intrinsic hole doping and the increase in resistivity for high doping provides insights into the consequences of $SF_6$ etching on graphene. In a broad sense, these observations tell us inhomogeneity is being introduced. We note the hole doping evident from the quantum transport measurements – 0.1 x $10^{12}$ cm$^{-2}$ is consistent with our room temperature Raman measurements, suggesting Raman spectroscopy is an excellent tool to track the etching process and quality of the underlying graphene layer. This origin of the doping inhomogeneity could be due to some slight fluorination which is known to dope graphene[7], or may potentially originate from molecular absorbates and/or polymer contamination[8]. However, our Raman spectroscopy measurements showed that after AFM brooming, the etched region returned to its pristine form. This suggests the mobility degradation originates from surface contaminants which can be removed through AFM brooming.

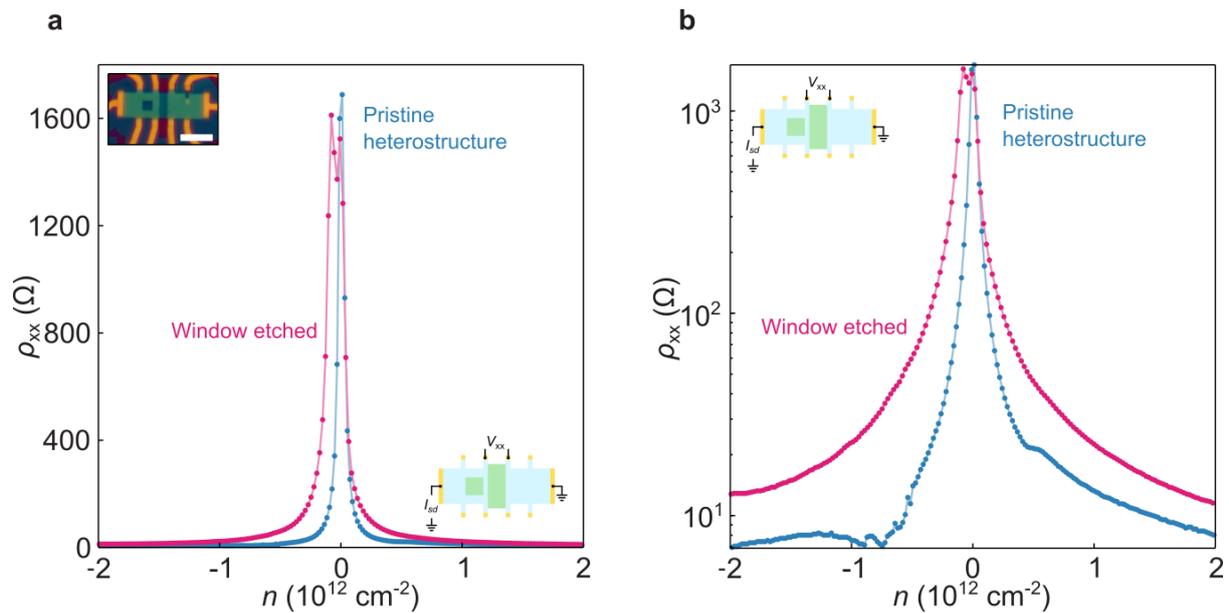

**Figure. S5 a,b** Quantum transport measurements of the longitudinal resistivity. **a,** Resistivity ($\rho_{xx}$) plotted as a function of carrier doping (*n*). The blue and pink colours plot data before and after etching respectively. Top left inset is an optical image of the device with scale bar of 10 μm. Bottom right inset depicts the measurement schematics, with green areas highlighting exposed graphene regions. **b,** same as **a** but plotted on a logarithmic scale.

**Supplementary Section 6 – Effects of over-etching**

Our experiments presented in D1_window1/2 and D3_window4 were made using the sequential etching process necessary for exposing clean open surfaces without degrading graphene's electronic quality. Here we present the consequence of over-etching showing how slight fluorination can degrade graphene´s quality. Fig. S6a plots the Raman maps for the pristine, etched and etched region post annealing. It shows with over exposure strong doping and strain changes are observed. Our analysis suggests doping inhomogeneities of around 1 x $10^{12}$ cm$^{-2}$. Fig. S6b plots low-temperature quantum transport measurements performed in the same device. The pristine region shows width limited mobility (Fig. S6c) and the usual gate transfer characteristics of high-quality encapsulated graphene. After over etching, we notice the quality degrades significantly (orange curves in Fig. S6b). We find a broad peak emerges at carrier doping around -1 x $10^{12}$ cm$^{-2}$. This is consistent with the observation of in Raman analysis (Fig. S6a) which shows carrier doping of similar magnitude. Notably, this over-etching results in a degradation of the carrier mobility by an order of magnitude. Understanding the origin of this degradation is non-trivial. While fluorination will degrade mobility, it should be detectable via D' peaks in our Raman measurements. Moreover, fluorination it is expected to hole dope the system[7], while contrasts with our transport data that shows a strong extrinsic electron doping (Fig. S6b). Whatever the origins of the degradation, our data shows the etching recipe must be calibrated carefully before opening windows in the devices. We note however, after annealing the sample exhibits some additional changes in the Raman response (Fig. S6a) suggesting fluorination may be reversible under suitable annealing conditions.

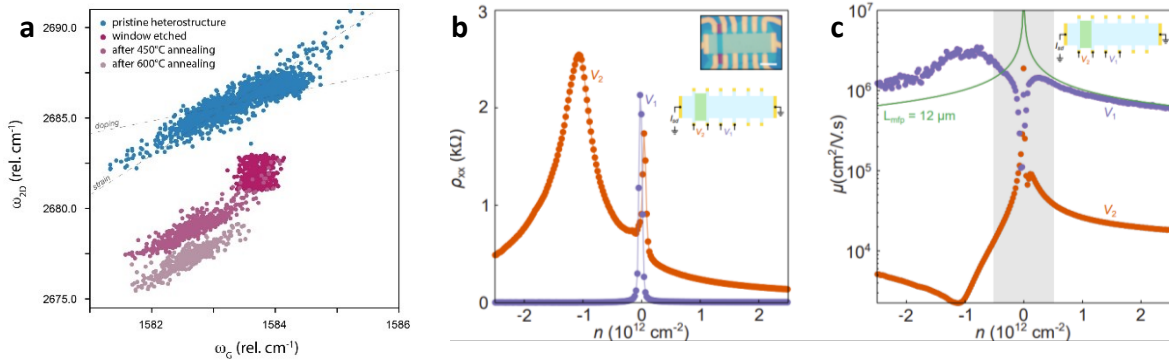

**Figure. S6 – over-etched devices. a,** Raman spectroscopy measurements plot the position of the 2D peak ($\omega_{2D}$) as a function of the G peak ($\omega_{2G}$) for different spatial positions in the pristine (blue data) and etched (red) regions. The different shades of pink plot measurements performed directly after etching, and after different annealing treatments. **b,** Resistivity ($\rho_{xx}$) as a function of carrier doping ($n$) plotted before (purple) and after (orange) etching. Insets plot the optical image after etching and the measurement schematic. Measurements are performed at 10 K **c,** carrier mobility plotted as a function of carrier density ($n$) for the data set presented in **b**. The green solid-line plots the theoretical width limited mobility.

**Supplementary Section 7 – Annealing**

As discussed in Supplementary Section 6, we found a large degradation in carrier mobility in D2_window3 after etching. The Raman combined with quantum transport data suggests the origin lies in adsorbates or contaminants rather than excess fluorination. This was because we do not see the emergence of a D' peaks in the Raman spectrum neither hole doping in the quantum transport measurement that typically accompanies fluorination. Here we present AFM measurements on D2_window3 (inset of Fig. S7a) to further corroborate these claims. Figure. S7a plots a spatial map of the height profile ($t$) corresponding to a 2 x 2 µm². We find sharp jumps around 11 nm scattered throughout the scan area we attribute to polymer contaminants on the surface. This may originate from the PMMA etch mask which was not properly broomed before etching. Figure. S7b plots the same spatial map after annealing at 600° C. We find that the polymer spikes are completely removed and the surface roughness decreases significantly in the sub nanometer range. This shows that high-temperature annealing can be used as a post-etching treatment to obtain smooth open surfaces.

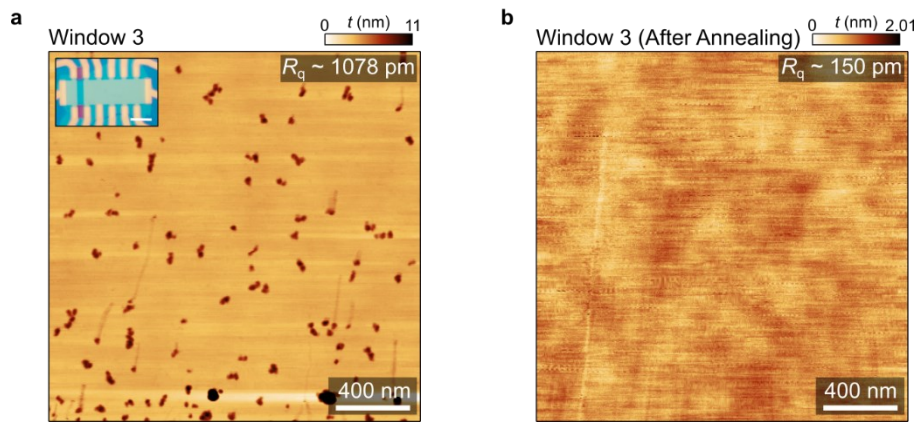

**Figure. S7. a,** AFM image of D2_window3 after etching. Inset: optical image of the device. Scale bar in main panel is 400 nm. **b,** same as **a**, after annealing at 600 °C.

**Supplementary Section 8 – selective etching of WS$_2$/graphene/WS$_2$ graphene heterostructures**

In the main text we showcase the potential of our selective etching technique for opening windows in graphene encapsulated with hBN devices. Here, we demonstrate its applicability to other 2D materials, specifically Tungsten Disulfide (WS$_2$). Figure. S8a plots a schematic of the heterostructure composed of WS$_2$/Gr/WS$_2$ on silicon dioxide (SiO$_2$). The optical image of the stack is presented in Fig. S8b. We use a similar recipe to that presented in Supplementary Section 1 recalibrated for working the WS$_2$. Using the same pre-etching steps, we etched a window on the device (black dashed box in Fig. S8b). To characterize the channel, we performed Raman spectroscopy measurements comparing the full width half maxima (FWHM) of the 2D peak (Fig. S8c). The spectrum for a fixed position (green cross in Fig. S8b) is plotted in Fig. S8e. The spectra naturally contrast with hBN encapsulated devices lacking the hBN phonon modes close to the D peak. Importantly, even after etching no pronounced D peak was observed confirming the fluorination of graphene was minimal. To benchmark how the quality of graphene changes during etching, we make spatial maps of the FWHM of the 2D peak (Fig. S8c). Notably, it shows the similarly low values within fully encapsulated and exposed regions suggesting the quality of graphene is preserved.

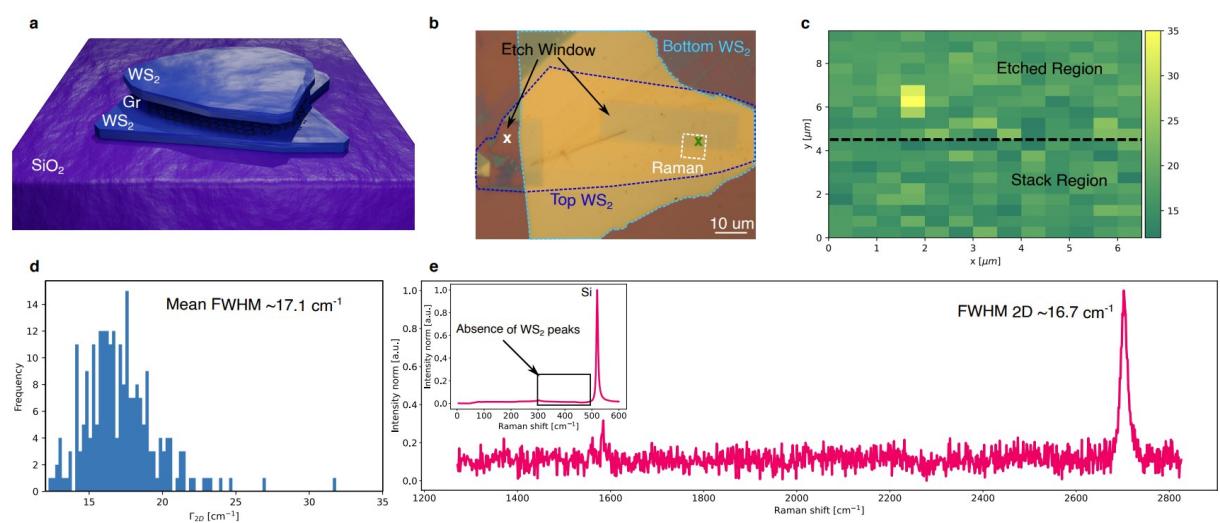

**Figure S8 Window etching in WS2/Gr/WS2 heterostructures. a,** Schematic of the heterostructure studied consisting of graphene (Gr) encapsulated in tungsten disulfide (WS$_2$). **b,** Optical image of the stack deposited on a silicon/silicon dioxide substrate (SiO$_2$). The light blue shading outlines the window region that was exposed using SF$_6$. Black arrows point to the etched window region and a control region (white cross) to verify the complete removal of WS$_2$. **c,** graph plots a spatial map (white dashed box in **b**) of the full width half maximum of the 2D peak of graphene ($\Gamma_{2D}$). The black dashed line indicates the interface between fully encapsulated and window regions. **d,** histogram of $\Gamma_{2D}$ for the map in **c**. **e,** Raman spectrum taken in the window region (green cross in **b**). inset plots a spectrum taken outside of the heterostructure in the wavelength range where WS$_2$ peaks appear. It shows only Silicon modes appear verifying the top WS$_2$ is completely removed.